# Star Formation in Galaxy Collisions: Dependence on Impact Velocity and Gas Mass of Galaxies in GADGET-4 Simulations

Authors: Gustavo Neves Pereira[1], Paulo Laerte Natti[2*]

[1]Departamento de Física, Universidade Estadual de Londrina, Rodovia Celso Garcia Cid, PR-445, Campus Universitário, 86037-390, Londrina, Paraná, Brasil. E-mail: gustavonevespereira@gmail.com, ORCID https://orcid.org/0009-0001-2125-0223

[2]Departamento de Matemática, Universidade Estadual de Londrina, Rodovia Celso Garcia Cid, PR-445, Campus Universitário, 86037-390, Londrina, Paraná, Brasil. E-mail: plnatti@uel.br, ORCID https://orcid.org/0000-0002-5988-2621

*Author correspondent

Abstract:
This work investigates variations in the star formation rate during galaxy collisions when the initial conditions of velocity and gas mass are altered. For this purpose, hydrodynamic simulations were performed using the GADGET-4 code, with initial conditions generated by the Galstep and SnapshotJoiner programs. Systems of two galaxies on a head-on collision course were modeled with relative initial velocities ranging from 100 km/s to 1000 km/s, considering two scenarios: the first with identical galaxies, and the second with galaxies of different sizes. In simulations of systems with higher initial relative velocities, both found more intense peaks in the star formation rate, triggered by the first contact of the collision, followed by a strong decline caused by gas dispersion. In contrast, for systems with lower initial velocities, mergers between galaxies were observed, leading to multiple peaks in the star formation rate. A greater initial distance between galaxies has also been linked to whether or not the galaxy system merges, since it implies longer timescales for gravitational action, which leads to higher relative velocities at the moment of collision. Furthermore, the star formation rate in galaxies was found to have a clear dependence on initial gas content. Furthermore, the initial gas content in galaxies revealed a clear dependence on star formation rates. Overall, our results show that the relative impact velocity, the initial distance between the galaxies, and the gas content are important parameters for analyzing the star formation rate in colliding galaxies.

## 1 - Introduction

Galaxy interactions are among the causes of changes in the star formation rate of a galaxy. These changes can lead to star formation rates much higher than expected for a galaxy of comparable mass, as in the case of the galaxy M82, which exhibits a very high star formation rate, a state known as a starburst, caused by interaction with a neighboring galaxy [1].

Among the types of interaction between two galaxies, collisions stand out. In these collisions, changes occur in the rate of star formation due to the interaction of gases present in the disks of these galaxies. High-speed collisions compress the gas to high densities, contributing to increasing the rate of star formation [2] and, consequently, to the reduction of the gas content in the galaxy after the collision [3].

Variations in the star formation rate of galaxies due to collisions are influenced by several parameters. These include the relative velocity of the galaxies during the collision, the distance between the galactic nuclei, i.e., (this distance being in a direction different from the direction of the approach speed called the impact parameter), whether the collision is head-on or not, the relative size and the gas densities of the galaxies, among other parameters. These parameters affect both the star formation rate [2] and the amount and dynamics of the remaining gas after the interaction [2,4].

To perform this type of analysis of galaxy collisions, involving disks and gas clouds, three-dimensional hydrodynamical simulations are carried out using computational tools such as RAMSES [5], GADGET [6], and Arepo [7]. These codes model the dynamical evolution of galaxies based on their initial parameters, which can, in turn, be generated with the aid of tools such as Galstep [8] and SnapshotJoiner [9]. The results of these simulations can then be accessed through visualization and analysis software such as Glnemo2 [10] and Gadgetviewer [11], which process and extract data from the simulations for both qualitative and quantitative study.

In these numerical experiments, whose objective was to analyze changes in the star formation rate in a system of two colliding galaxies, the Galstep and SnapshotJoiner tools were used. Thus, a pair of standard and stable galaxies were generated and placed on a collision course. These initial conditions, after being executed in the GADGET-4 program, which was configured and parameterized to allow star formation from the gaseous disks, had their results visualized using the Gadgetviewer tool. Such visualizations enabled analyses and interpretations of the collision results.

This article is organized as follows. Section 2 describes the methodology for generating the initial conditions of galaxy composition and the parameters of galaxy collisions. Section 3 presents and discusses the simulation results, focusing on the behavior of the galaxy system's gaseous disk and the star formation rate. Section 4 analyzes these results and compares them with data from the literature.

## 2 - Materials and methods

In this work, the computational tool GADGET-4 and its libraries are used to simulate galaxy collisions.

### 2.1- Software GADGET-4

Originally developed by Volker Springel, Naoki Yoshida, and Simon D. M. White in the year 2000, GADGET (GAlaxies with Dark matter and Gas intEracT) is a tool for hydrodynamic and N-body cosmological simulations. It was initially released in two versions: one for single processors, being the serial version, and another for multiple processors, which makes it possible to divide the calculations among several processors to improve computational performance and carry out simulations with more particles (approximately 75 million particles in this version) [16].

In 2005, GADGET-2 was released, featuring some technical improvements over its predecessor, such as the use of multiple time steps for calculating particle interactions in regions of different densities and accelerations. These allowed for better use of multiple processors, aiming to avoid bottlenecks and to achieve better conservation of the system's energy through calculations involving entropy as a fundamental variable. This version was capable of simulating more than 250 million particles and analyzing both radiative cooling and star formation in these systems [17].

While GADGET-3 was not made publicly available, GADGET-4 was released in 2020, bringing improvements to the features introduced in GADGET-2, such as time operations and parallelization across multiple processors, as well as a new pressure-based approach in hydrodynamical simulations, while still maintaining compatibility with older models [24].

Its operation consists of using an initial condition, which contains all the initial data of the analyzed system, such as the masses, velocities, and positions of each particle, a condition that is usually generated by another program. This initial condition and the other simulation parameters (such as gravitational constants, precision settings, star formation parameters, and runtime options) are all specified in a text file that is then executed by the program, producing several snapshots with data on the evolution of the system, along with text files providing specific data on other quantities during the system's evolution [6].

**2.2 – Software Glnemo2**

The Glnemo2 program is an interactive visualization software for cosmological simulations produced for programs such as GADGET. When loaded with the output of a simulation, it allows the particles to be visualized, making it possible to obtain data on the structure of galactic systems, the formation of halos, filaments, and arms, among others. It also provides data specifically for the particles of each type in the simulation, whether they belong to the halo, bulge, or disk. It is designed to handle millions of particles and allows separate analysis of subsets of these systems [10, 20].

Given the level of detail and the number of particles, Glnemo2 relies largely on a dedicated GPU (Graphics Processing Unit), which makes its use difficult on machines that have an integrated GPU, as stated in its installation requirements [20, 21].

**2.3 – Software Gadgetviewer**

Similar to Glnemo2, the Gadgetviewer program is also an interactive visualization software for cosmological simulations produced for programs such as GADGET, which allows us to observe data on the structure of the simulated galaxy systems. Its operation consists of selecting and loading a snapshot generated by GADGET, which then enables the generation of a 3D model with the particles. The software also provides data on star formation, metallicity, mass, and densities, providing average, maximum, and minimum values for each quantity across various possible selections of particles in the snapshot. One can also analyze the temporal evolution and generate a video of this progress.

Unlike Glnemo2, Gadgetviewer is much less demanding and can be run on an integrated GPU [11]. Given this lower computational requirement for generating stored images, it was used in the simulations via a computer equipped with an integrated GPU.

**2.4 – Software Galstep**

The Galstep program generates galaxy initial conditions, which can then be used as input for simulations with codes such as GADGET. Its operation consists of choosing several initial parameters, such as mass, number of particles, size, and density of stellar disks and bulges, gaseous disks, and dark matter halos present in galaxies. A larger number of particles defined in the initial parameters provides greater precision in the simulations, since fewer approximations are made in the numerical calculations; on the other hand, these larger numbers also impose greater computational capabilities for carrying out the simulations [6, 8, 12].

After selecting each of these parameters, Galstep is executed, and an HDF5 file is generated containing the data of that galaxy under these initial conditions. This type of file is necessary for carrying out simulations in programs such as GADGET [6].

**2.5 – Software SnapshotJoiner**

Galstep is therefore quite useful for creating the initial conditions of a single galaxy. However, for the study of collisions, an initial condition file containing both galaxies on a collision course is required. To generate this second file, the **SnapshotJoiner** is used. Its operation consists of selecting two initial galaxy files, produced by programs such as Galstep, along with the relative position, velocity, and rotation conditions of these galaxies. Note that a 180° rotation around the Z-axis would make them "mirrored". In this context, a positive relative velocity indicates that the galaxies are moving apart, while a negative velocity indicates that they are approaching [9]. After selecting each parameter, the SnapshotJoiner is executed, and the inputs are compiled, generating an HDF5 file that contains the data of this two-galaxy system with the chosen initial conditions and relative positions/velocities. This file can then be used to analyze the evolution of the system in programs such as GADGET [6].

# 3 - Results

### 3.1 – Initial Conditions

To obtain the modeling data of a system of two galaxies in a frontal collision, the program Galstep is used, which generates the initial conditions of each galaxy from parameters provided for the halo, the disk, the bulge, and the gas in the galaxy. For the simulations carried out in this work, two standard galaxies were created.

For the first galaxy generated by Galstep, designated as the large galaxy, the initial ratio of gas to other types of baryonic matter was set at 14%, therefore a mass of $5 \times 10^{10} M_\odot$ was assigned to the disk, $10^{10} M_\odot$ to the bulge, and $10^{10} M_\odot$ to the gas disk (in this way, adding the baryonic components, a total mass of $7 \times 10^{10} M_\odot$ is obtained, whose gas part, $10^{10} M_\odot$, is 14%). As for the halo, composed of dark matter, it was initialized with $10^{12} M_\odot$.

For the second galaxy generated by Galstep, designated as the small galaxy, it was initialized with a halo with half the mass, as well as the stellar disk, the gaseous disk, and the bulge. The radial dimensions were also reduced by a factor $\sqrt{2}$, to keep the gas and matter densities equal in both modeled galaxies.

An important part of this initial parametrization is also determining how this mass is distributed in the galaxies. Note that a larger number of particles provides a more accurate result, as it more closely approaches the number of particles present in real galaxies, although at the expense of a higher computational cost.

For the simulations performed with the large galaxy, the halo component, with an initial mass of $10^{12} M_\odot$, had its mass distributed into 10,000 particles, each of which is composed of large groupings of solar masses that make the simulation feasible [18]. In the same way, the disk was divided into 10,000 particles, the bulge into 4,000 particles due to its lower mass, and, as it is the focus of the experiment, the gas component was divided into 40,000 particles. The values for the radial scale factors for the disk and the gas was 3.5kpc, with a radial cut at 30kpc. The bulge and the halo follow the Denhen profile, with $\gamma = 1$ and a scale radius of 47kpc for the halo and 1.5kpc for the bulge, with the cuts at 900kpc and 30kpc respectively [8].

For the simulations performed with the small galaxy, all these numbers were reduced by half. All other Galstep parameters were kept as the default. Finally, the program was executed to create the two galaxy models.

Once these galaxy models were created, the SnapshotJoiner program was then used, which is designed to combine two galaxies into a single system, simulating a frontal collision, for example. For this, it was necessary to define two important parameters: the initial distance between the galaxies and their initial relative velocity. For these experiments, the relative velocity between the galaxies was varied from 100 km/s to 1000 km/s, in increments of 100 km/s, considering that the initial distance between the galaxies was 200 kiloparsecs. Given these parameters, ten galaxy systems consisting of two identical galaxies (two large galaxies) simulating a frontal collision were constructed. Similarly, ten other galaxy systems were constructed, each composed of two non-identical galaxies (one large galaxy and one small galaxy), also simulating a head-on collision.

After constructing the initial conditions of these two sets of colliding galaxy systems, the simulations could be performed with the GADGET-4 program. In these simulations, the STARFORMATION function [6] was enabled, which allowed the formation of stars from the gas present in the gaseous disk of the galaxies.

## 3.2 – Numerical simulations: GADGET-4

With GADGET-4 properly configured to describe the frontal collision of a galaxy system, the simulations were executed. Each simulation resulted in several snapshot files, each containing a moment of the simulated collision between the galaxies. Each snapshot was analyzed by the Gadgetviewer program, generating a visual interface for the positions of the particles in the simulations at each snapshot, making it possible to obtain information about the dynamics of the galaxy collision, including the generation of videos.

In addition, another result from GADGET-4 was analyzed. The file sfr.txt, generated in each of the simulations [6], provides the total star formation rate (in solar masses per gigayear). The successive data extracted from the sfr.txt file have a time interval of approximately $1.5 \times 10^{-4}$ Gyr. The graphs of star formation rate, as a function of time, were plotted using Microsoft Excel.

### 3.2.1 – Collision of two large galaxies as a function of relative speed

In these ten simulations, the two large galaxies analyzed started with the same relative distance, the same masses, and particle numbers. Thus, the first snapshot generated by GADGET-4 for each of the ten simulations shows the same initial configuration (Figure 1).

Figure 1 – Initial configuration of two large galaxies in the Gadgetviewer visual interface, with a relative distance of 200 kiloparsecs.

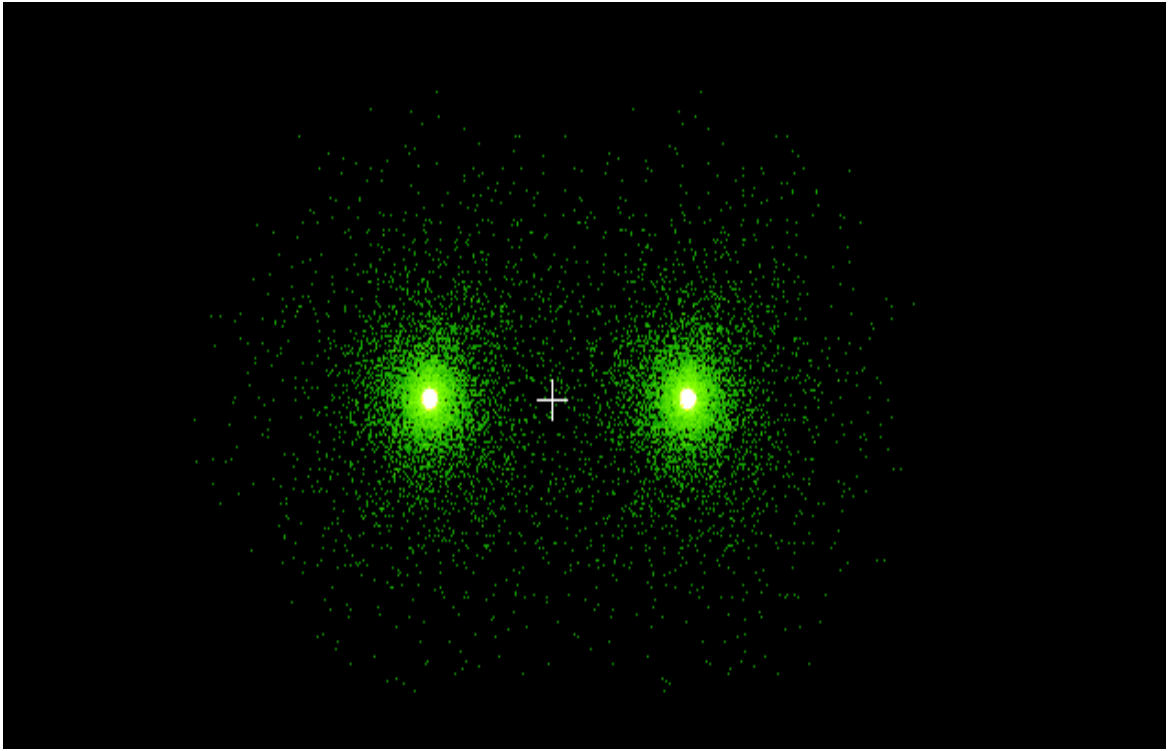

In Figure 1, the green points surrounding the galaxies correspond to dark matter particles, which interact only gravitationally. The white points at the centers of the galaxies are particles composed of baryonic components, that is, clusters of gas and stars present in the disk and bulge. The STARFORMATION parameter converts the gas in these clusters into stars during the collisions.

In the simulation presented in Figure 2, the galaxy system approaches with an initial relative velocity of 1000 km/s. After the collision, a large amount of gas (red points) is stripped and dispersed by the collision, and the galaxies proceed along opposite trajectories.

Figure 2 – Final configuration of the numerical simulation of a head-on collision between two large galaxies with an initial relative velocity of 1000 km/s.

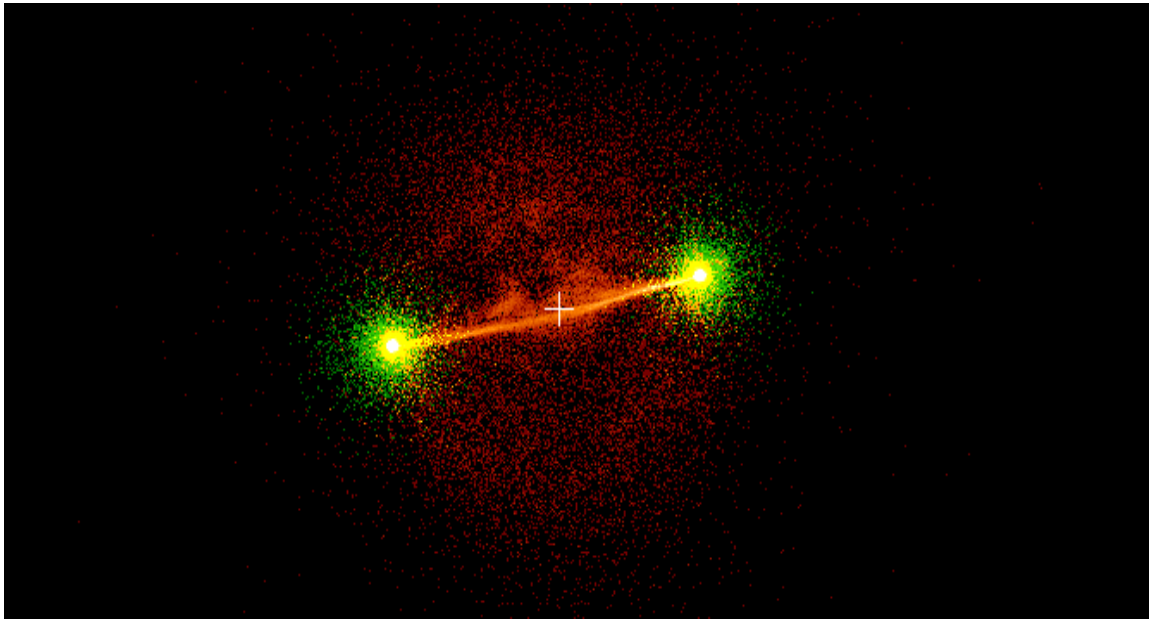

In Figure 2, the red points between the galaxies correspond to gas clusters. It is notable that after the collision, several of these gas clusters that were originally located in the galactic center were dispersed throughout the galaxy system [2]. Figure 3 is constructed from the sfr.txt files, which provide the star formation rate as a function of time in this simulation.

Figure 3 - Rate of star formation in the head-on collision of two large galaxies with an initial relative speed of 1000 km/s

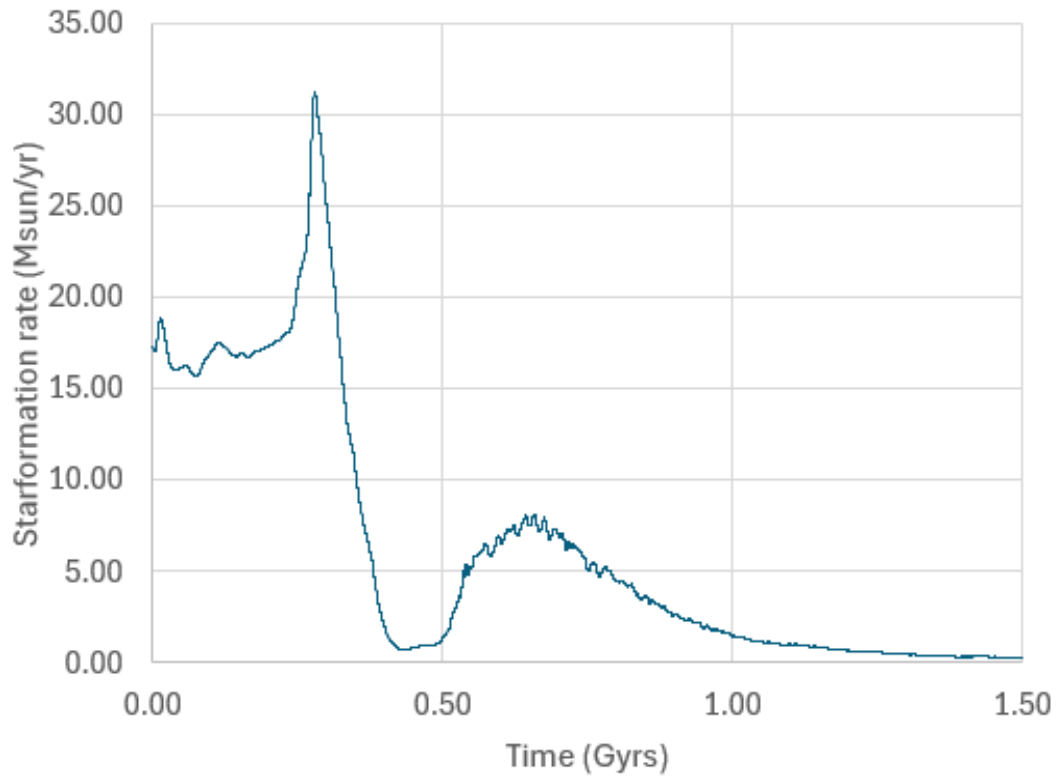

Numerical simulations for other initial relative velocity conditions were carried out. Figure 4 presents the star formation rates as a function of time for head-on collisions between two large galaxies, considering initial relative velocities from 100 km/s to 1000 km/s, in increments of 100 km/s.

Figure 4 - Rate of star formation in head-on collisions of two large galaxies with initial relative velocities ranging from 100 km/s to 1000km/s.

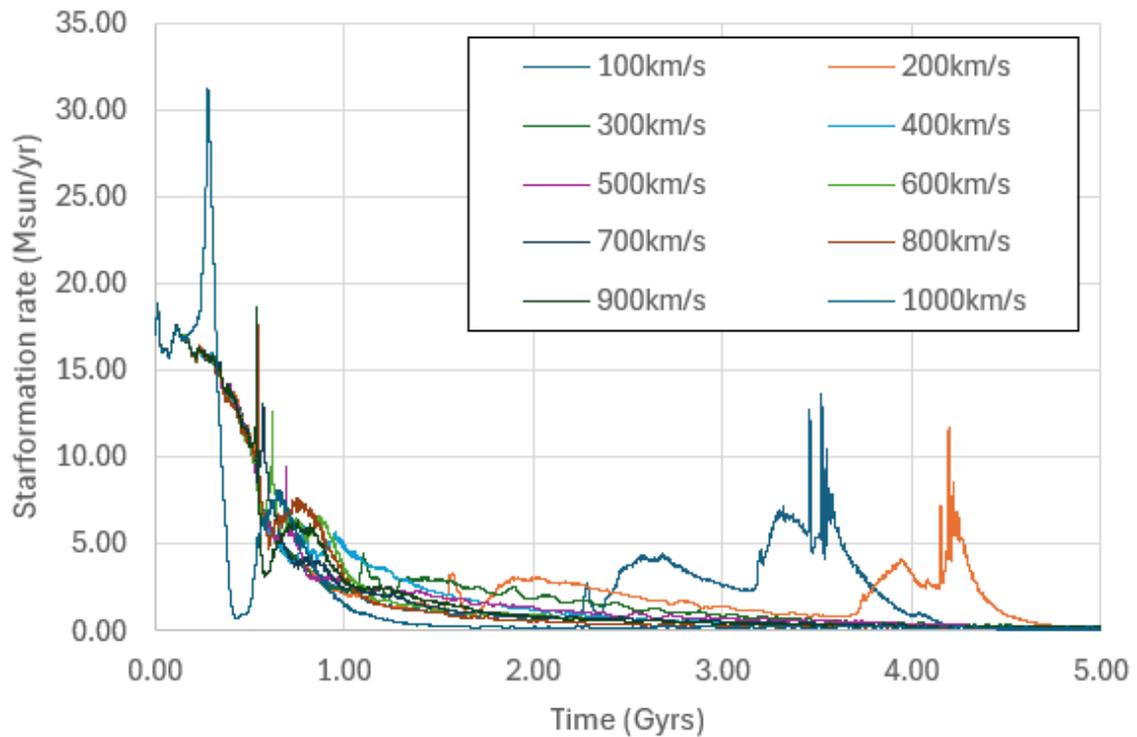

Figure 5 - Rate of star formation in head-on collisions of two large identical galaxies without merging.

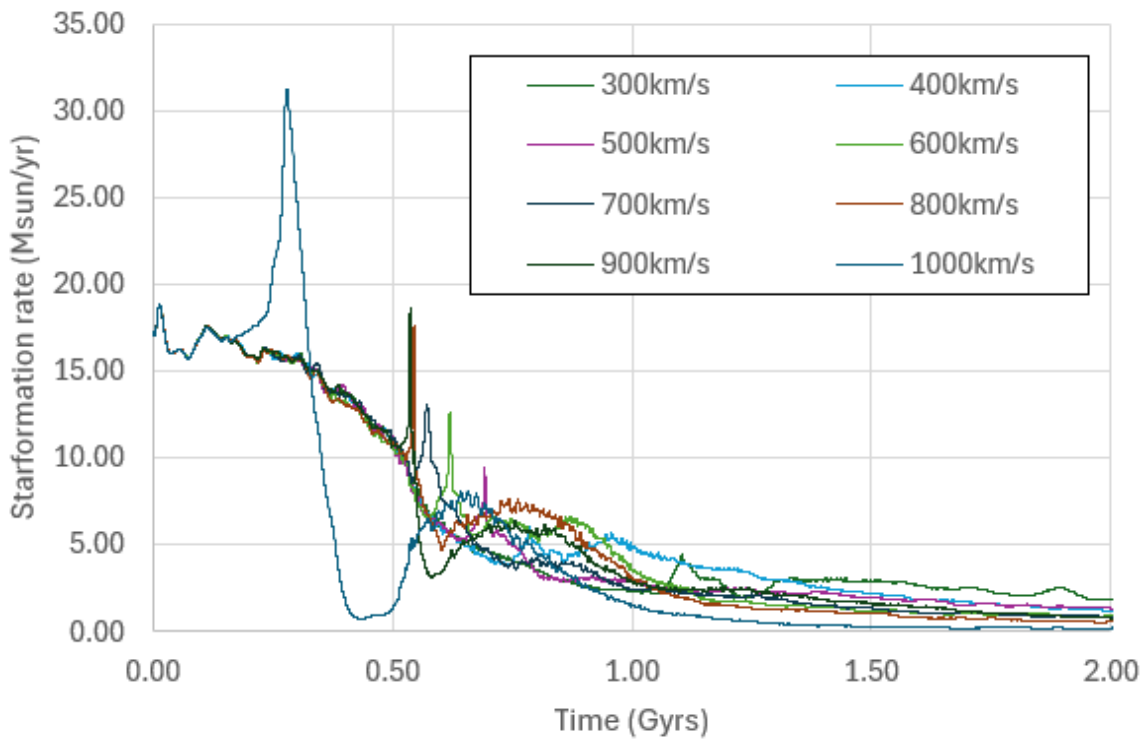

Figure 4 provides an overview of the star formation rate in head-on collisions of two large galaxies at different initial relative velocities. In collisions with higher relative velocities, the star formation rate exhibits behavior similar to that shown in Figure 3: some fluctuations in the star formation rate of the galaxy systems are observed before the collision, a sharply defined peak in the formation rate during the collision, followed by a steep decline, and then a return to a moderately decreasing star formation rate. The simulations that display this pattern are presented in Figure 5. In these simulations, after the head-on collision, the galaxies move apart in opposite directions without merging, similarly to the configuration shown in Figure 2.

Note in Figure 5 that in all simulations, during the head-on collision of two large identical galaxies, an initial peak occurs, whose timing varies with the initial relative velocity. After this peak, the star formation rate undergoes a steep decline, followed by a return to a gradually decreasing state, similar to the behavior observed before the collision.

On the other hand, there is another group of curves in Figure 4 corresponding to the cases in which the final configuration of the galaxy collision results in a single galaxy, that is, the galaxies undergo a merger, as shown in Figure 6.

Figure 6 – Final configuration of the collision of two large galaxies in which the merger process occurs.

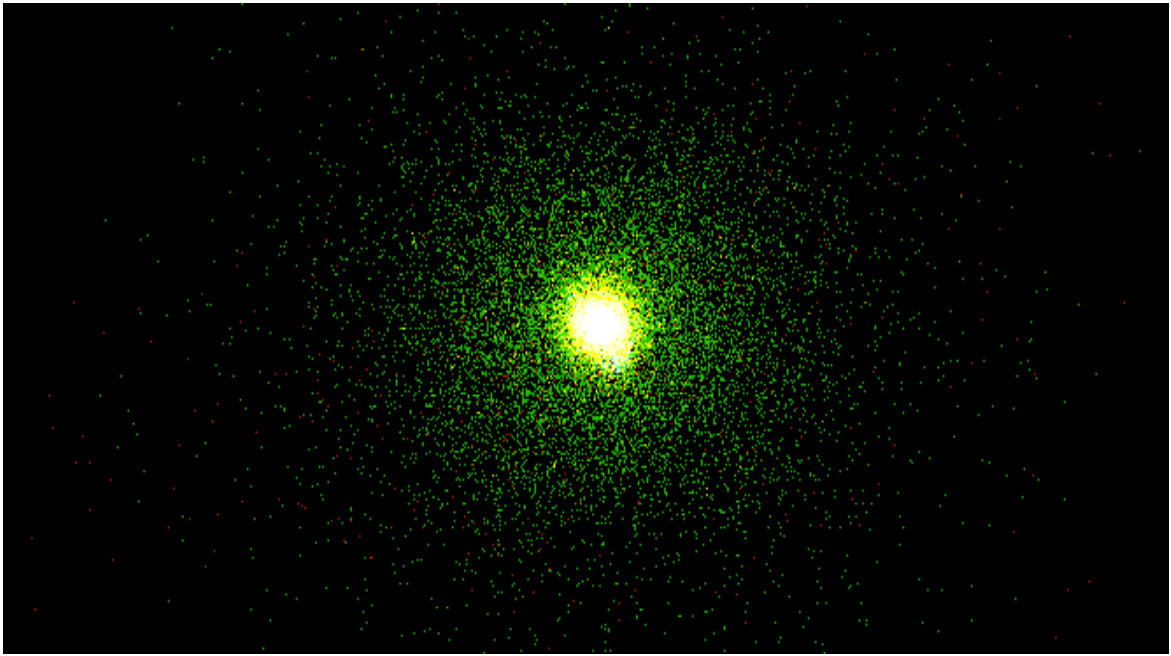

Figure 7 shows, separately, the star formation rates for head-on collisions of two large galaxies when a merger process occurs. In this process of galaxy merger, multiple peaks in the star formation rate arise as galaxies interpenetrate (secondary head-on encounters) in an oscillatory process until the merger process finally stabilizes.

The initial distance between galaxies is another important parameter for whether or not the merger process occurs. Considering the system of two large galaxies, with an initial relative velocity of 300 km/s, but with an initial distance of only 100 kiloparsecs, a merger process is observed between the galaxies, as shown in Figure 8, compared to the system with twice the initial distance. Note that, although this system shares the same initial relative velocities with one of the systems shown in Figure 5, that being the one with the initial relative speed of 300km/s, due to gravitational attraction and the different distances traveled, the relative velocities of the galaxies at the moment of the head-on collision are different, which explains the different final configurations of the galaxy systems.

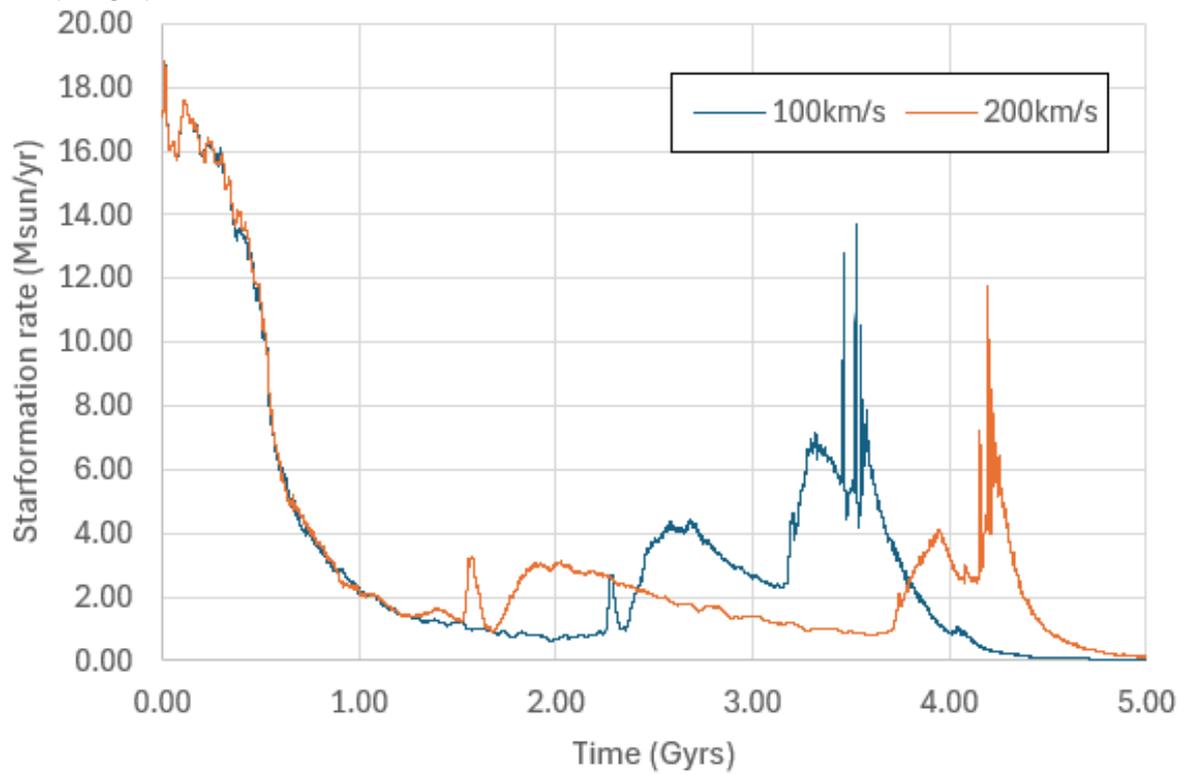

Figure 7 - Star formation rate of identical galaxy systems in head-on collisions with more than one peak (merger).

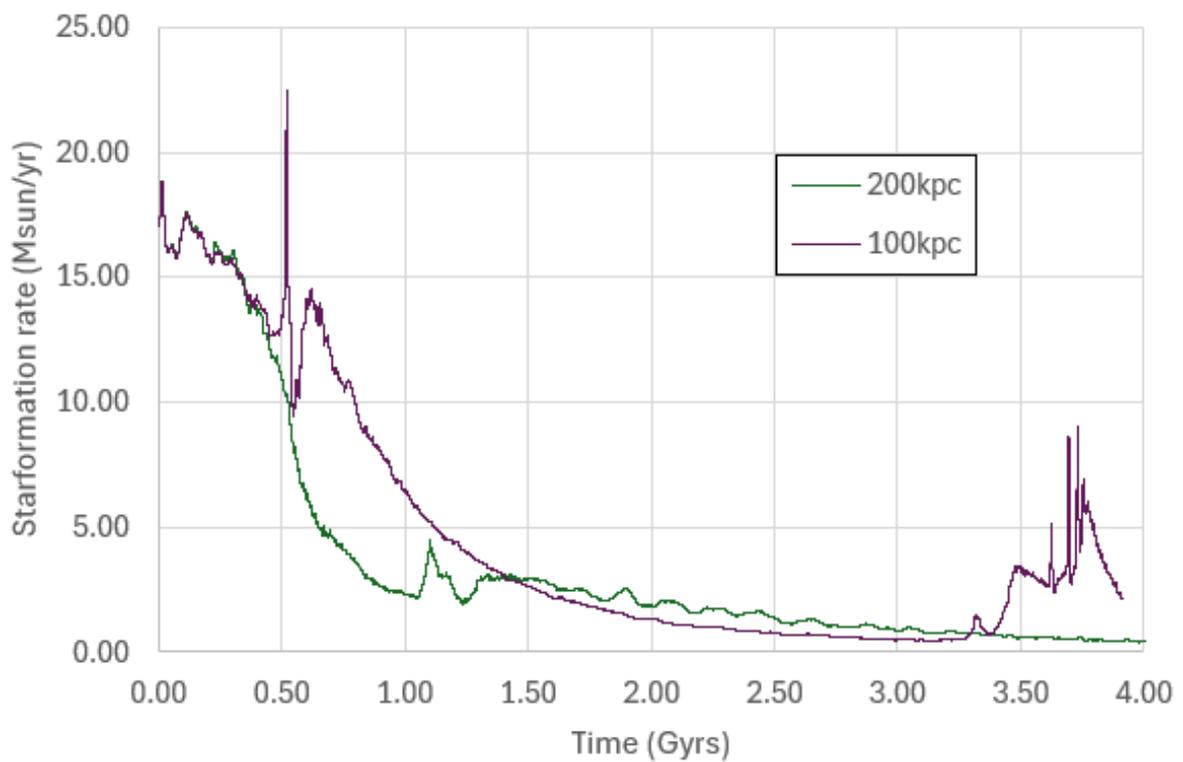

Figure 8 - Star formation rate for the system of two large galaxies in head-on collisions with an initial relative speed of 300km/s and an initial distance of 100kpc.

In Figure 8, in the system with the starting distance of 100kpc, an initial peak is observed, due to the first head-on collision, but after a long period, new peaks are observed. This dynamics of the star formation rate is characteristic of galaxy collisions involving a merger, with a final configuration similar to that shown in Figure 6, instead of the final configuration observed in the system with the starting distance of 200kpc, which is similar to Figure 2.

### 3.2.2 – Collision of large-small galaxies as a function of relative speed

In these ten simulations, the galaxies analyzed started with the same starting relative distance, but they have different masses and different numbers of particles. One of the galaxy is √2 times larger than the other, but they were modeled so that their gas densities are equal in both. The same applies to the matter densities. Finally, once again, the galaxies were initially placed 200 kiloparsecs apart. Thus, when examining the first snapshot generated by GADGET-4 for each of the ten different simulations, the same initial layout is obtained, as shown in Figure 9. In Figure 9, we see the small galaxy on the right and the large galaxy on the left.

As in the previous simulations, the green points surrounding the galaxies are particles representing dark matter, while the white points correspond to the baryonic disk and bulge components. Among these components there is gas, which can undergo star formation under the appropriate physical conditions, as prescribed by the parameters of the STARFORMATION subroutine.

Similar to the studies carried out for the head-on collision of two large galaxies, we first consider a configuration in which the initial relative velocity of the galaxies is 1000 km/s. Likewise, after the collision, there is a significant loss of gas (red points), with the galaxies moving in opposite directions, as shown in Figure 10.

Figure 9 – Initial configuration of large-small galaxies in the Gadgetviewer visual interface.

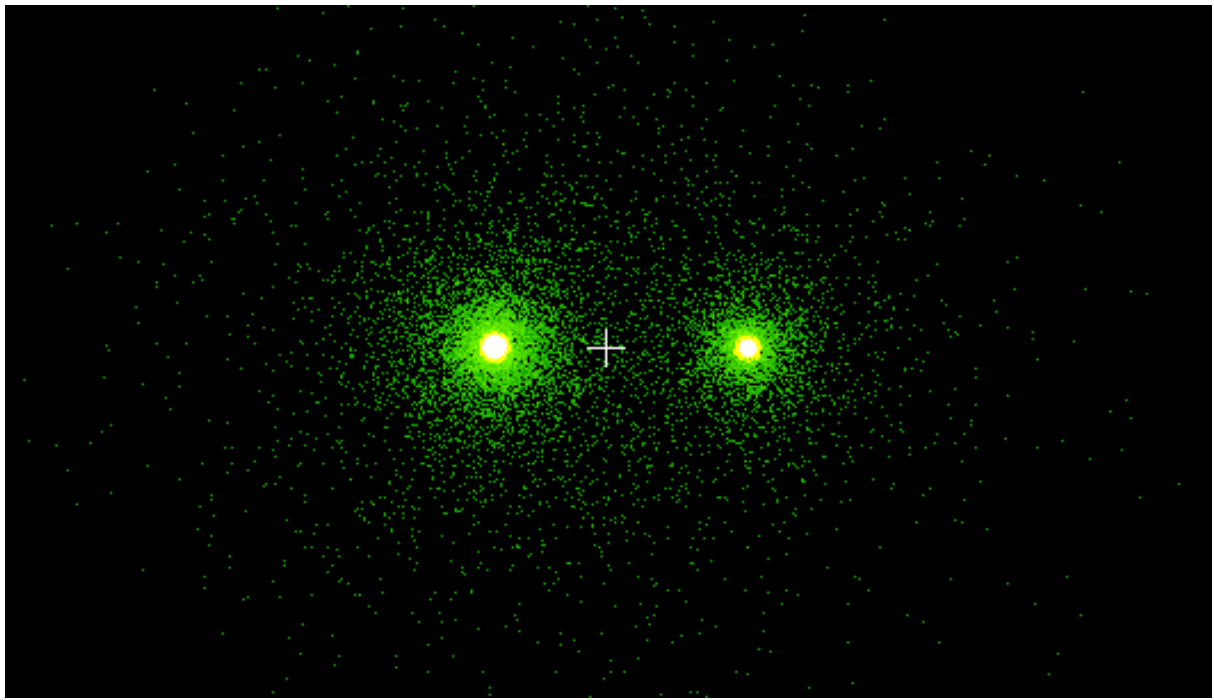

Figure 10 – Final configuration of the numerical simulation of the head-on collision of large-small galaxies with an initial relative speed of 1000 km/s.

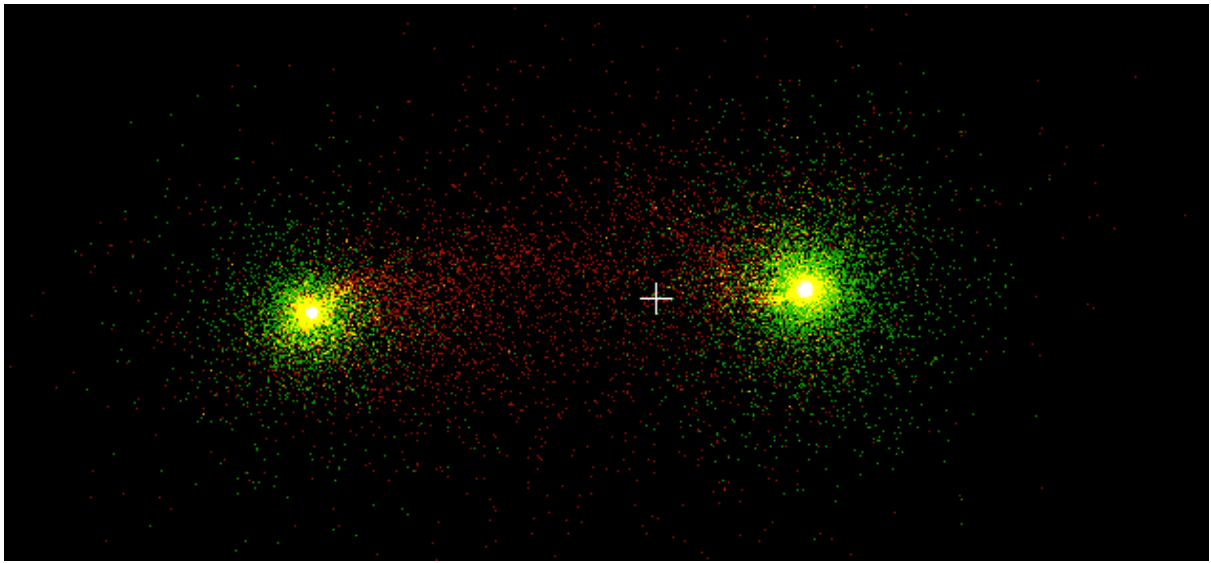

In Figure 10, after the collision, most of these gas clusters disperse into intergalactic space, while a portion is gravitationally attracted by the halo and falls back onto the galaxies [2].

Figure 11, constructed from the data provided in *sfr.txt*, displays the star formation rate as a function of time for this galaxy system.

Figure 11 - Star formation rate in the head-on collision of large-small galaxies with a relative speed of 1000km/s.

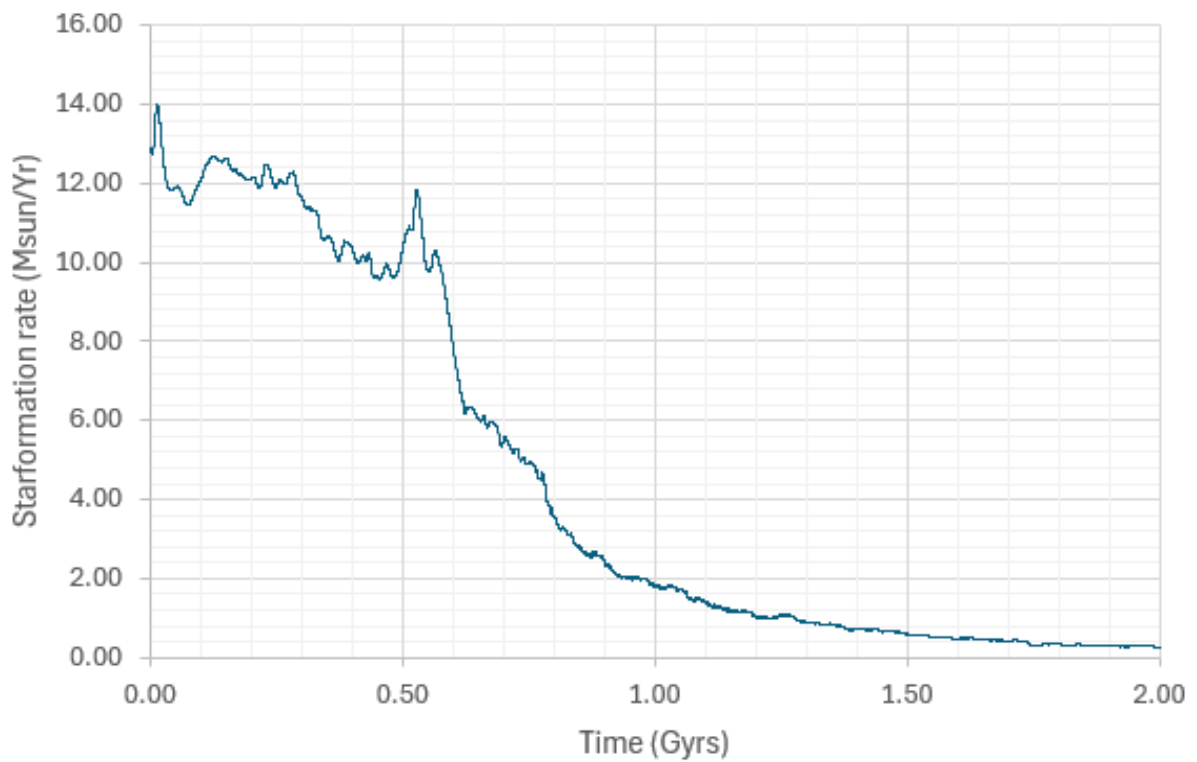

Comparing the star formation rates presented in Figures 3 and 11, it is found that with large-small galaxies, the timing of the collision differs, even when they start from the same initial distance. In Figure 3, the first peak is observed around 0.3 Gyr after the beginning of the simulation, whereas in Figure 11, this peak is observed at approximately 0.5 Gyr.

Another difference observed is the amplitude of these initial peaks, which in Figure 3 is much larger than the peak observed in Figure 11. Despite these differences, after the collision, both systems display the same behavior, with the star formation rate gradually decreasing.

Numerical simulations for other initial relative velocity conditions were carried out. In Figure 12, star formation rates as a function of time are presented for different initial relative velocities. Once again, the star formation rate curves correspond to relative velocities ranging from 100 km/s to 1000 km/s, in increments of 100 km/s.

Figure 12 – Rate of star formation in the head-on collision of large-small galaxies with initial relative velocities ranging from 100 km/s to 1000km/s.

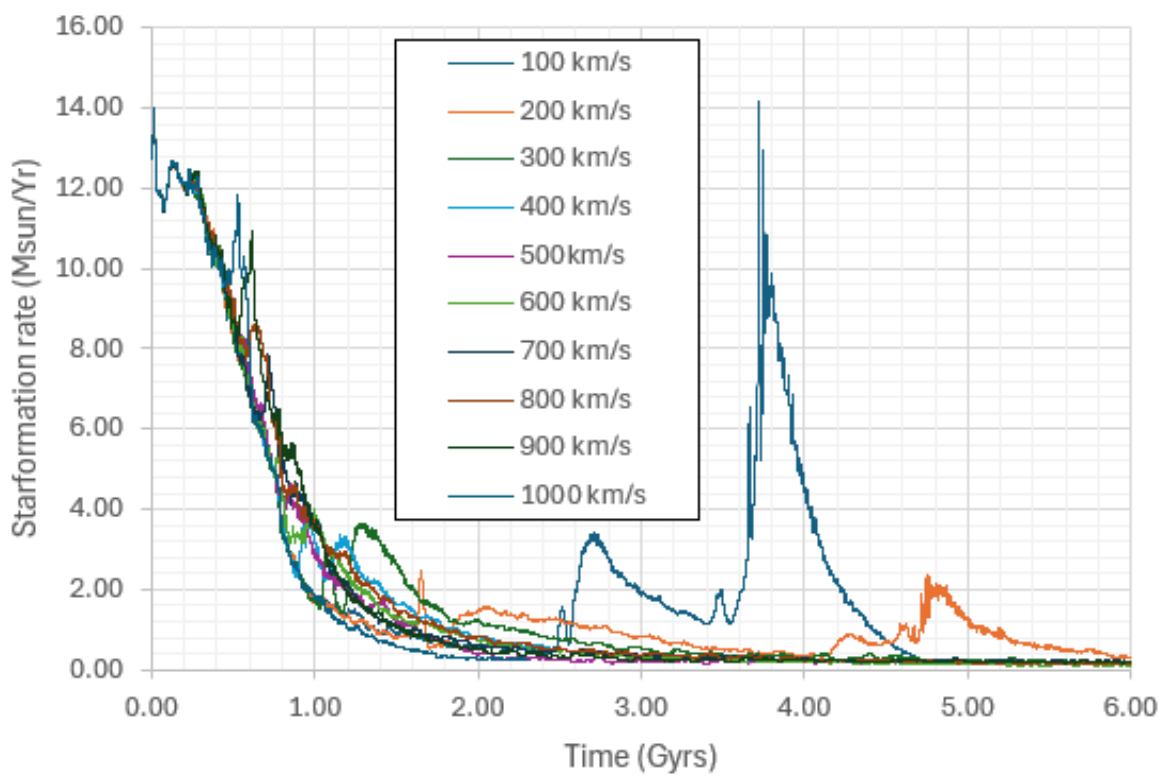

As in the case of the collision of two large galaxies, the numerical simulations can be classified into two categories: those that exhibit only one peak in the star formation rate, and those in which more than one peak is observed, a situation that occurs when galaxies merge. Figure 13 presents numerical simulations of galactic collisions, as a function of initial relative velocities, when a single peak is observed.

In Figures 11 and 13, some fluctuations in the star formation rate of the galaxy systems are observed before the collision, a sharply defined peak in the formation rate during the collision, and, after the impact, a gradual decline in the star formation rate, a behavior analogous to that previously observed in the collisions of two large identical galaxies.

In Figure 14, the star formation rates are presented for head-on collisions of large-small galaxies in which the merger process occurs. In this galaxy merger process, multiple peaks in the star formation rate are observed due to the oscillatory dynamics of the galaxies until they merge. Figure 15 presents the final configuration of these merging galaxy systems.

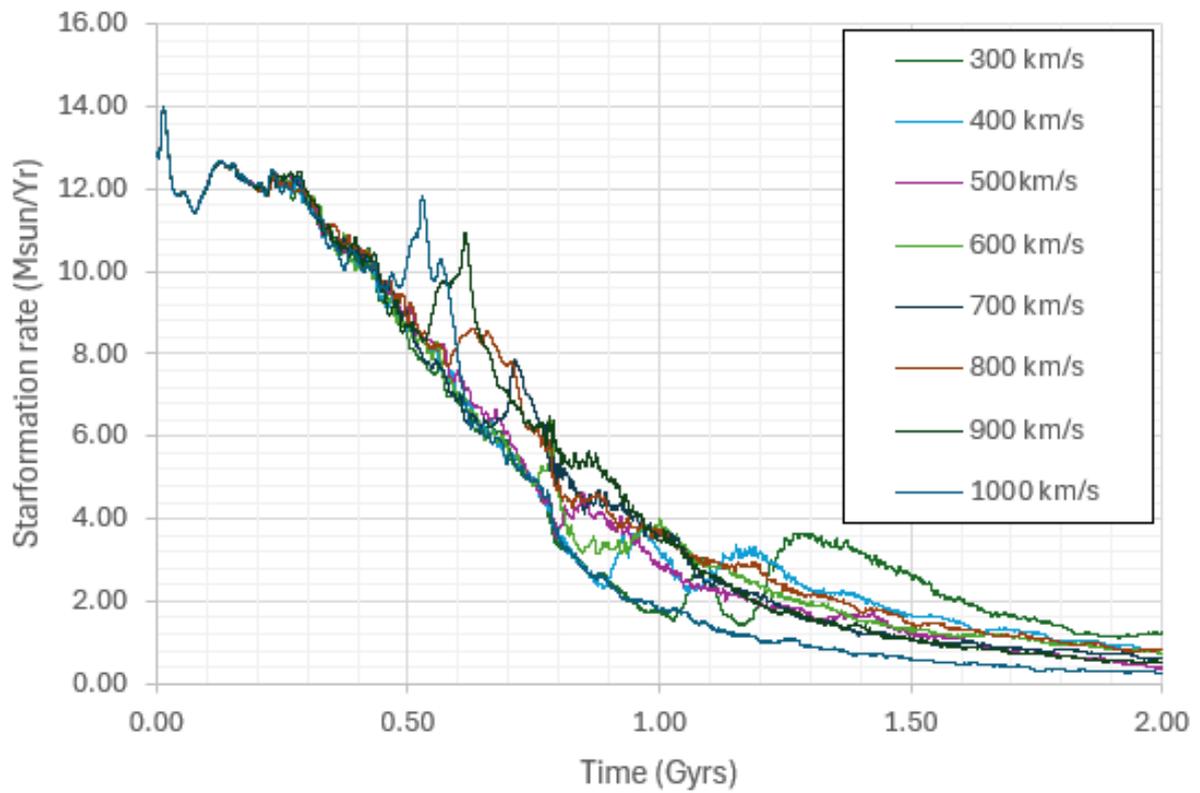

Figure 13 - Rate of star formation in head-on collisions of large-small galaxies without merging.

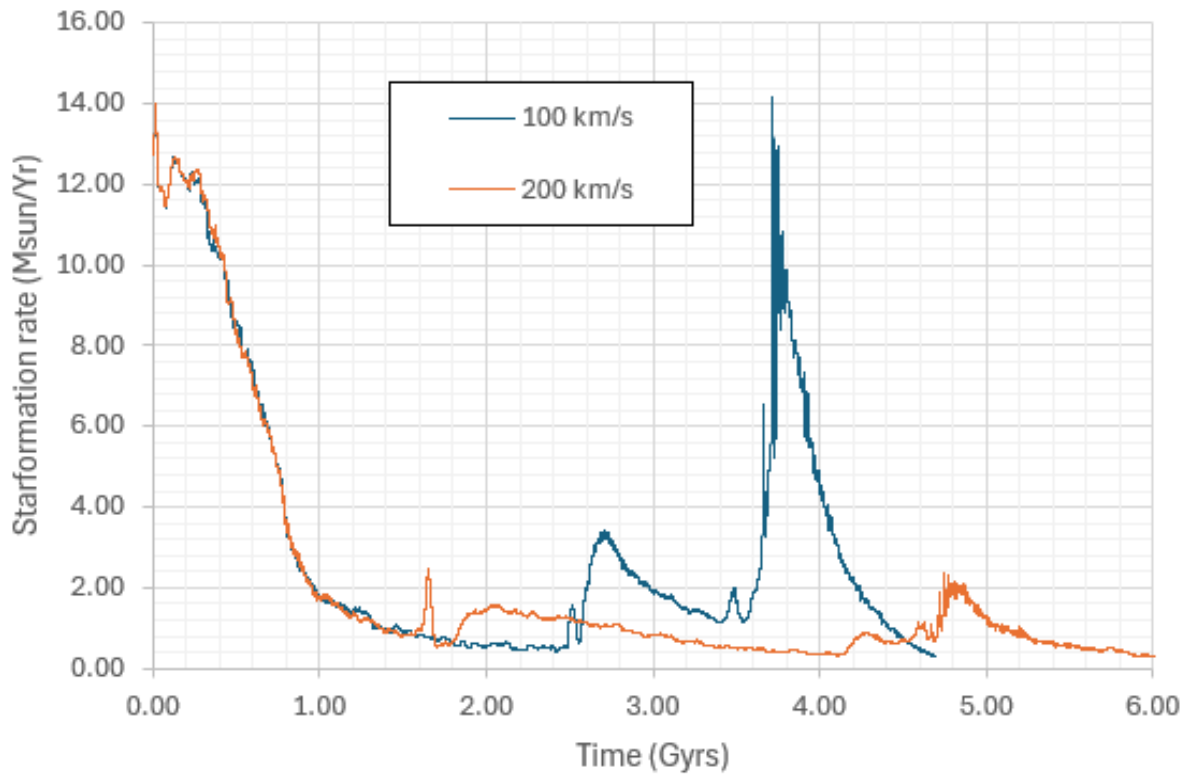

Figure 14 - Star formation rate of large-small galaxy systems in head-on collisions with more than one peak (merger).

Figure 15 – Final configuration of head-on collisions of large-small galaxies in which the merger process occurs.

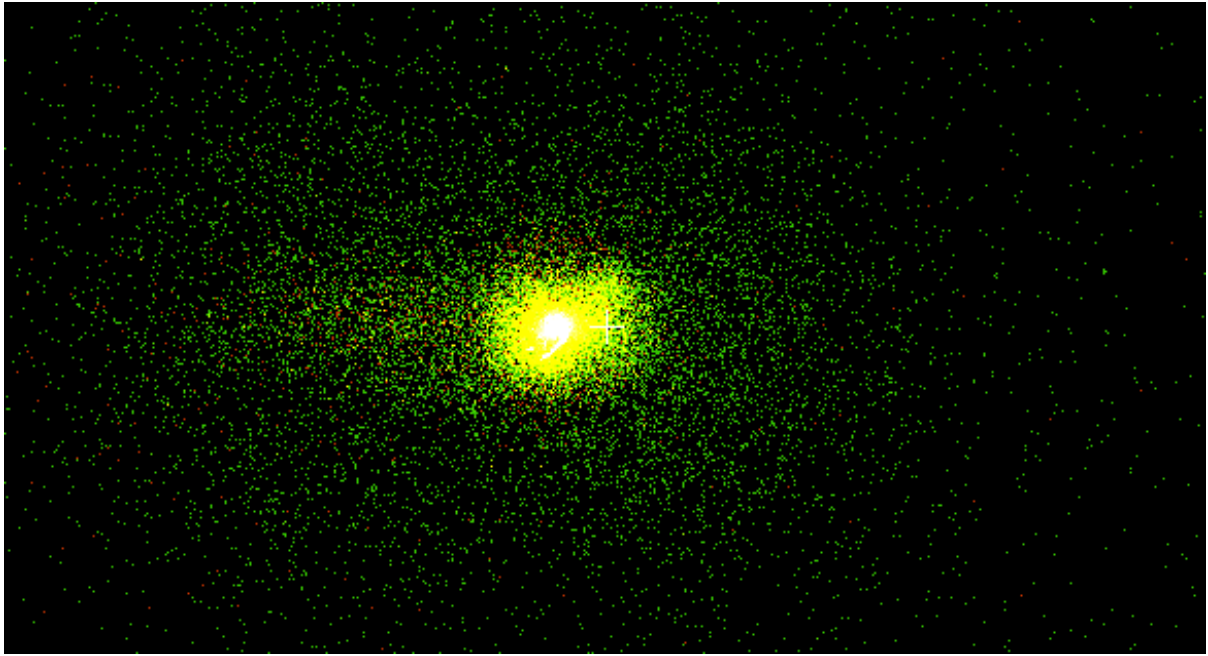

From the simulations performed, due to the greater amount of gas in the system of two large galaxies, the more massive galaxies exhibit higher star formation rates compared to the star formation rates in large-small galaxy systems. A comparison between the star formation rates in systems with identical and different masses, in cases where galaxy merger occurs, is presented in Figure 16.

Figure 16 – Comparison between star formation rates in head-on collisions of identical and different galaxies, when merger occurs.

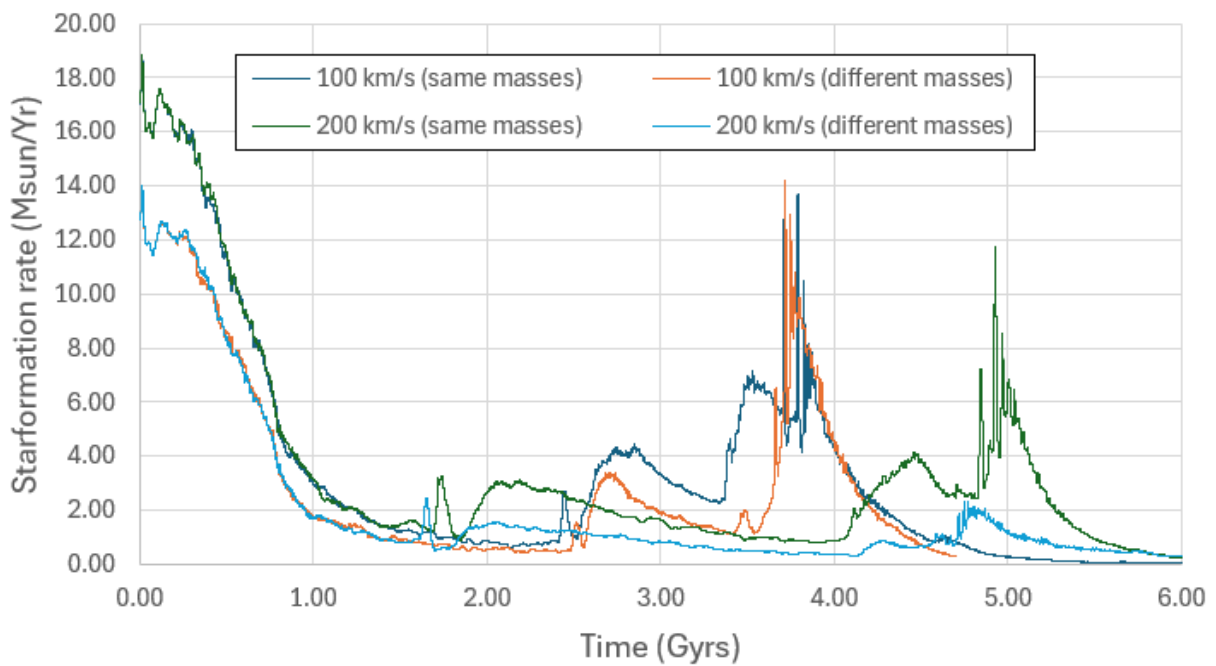

Similarly, it is observed that galaxy systems undergoing head-on collisions, when no merger occurs, also display higher star formation rates in the case of collisions between two large galaxies (more massive galaxies). It is also verified that, in such configurations, the peaks are higher, as can be seen in Figure 17.

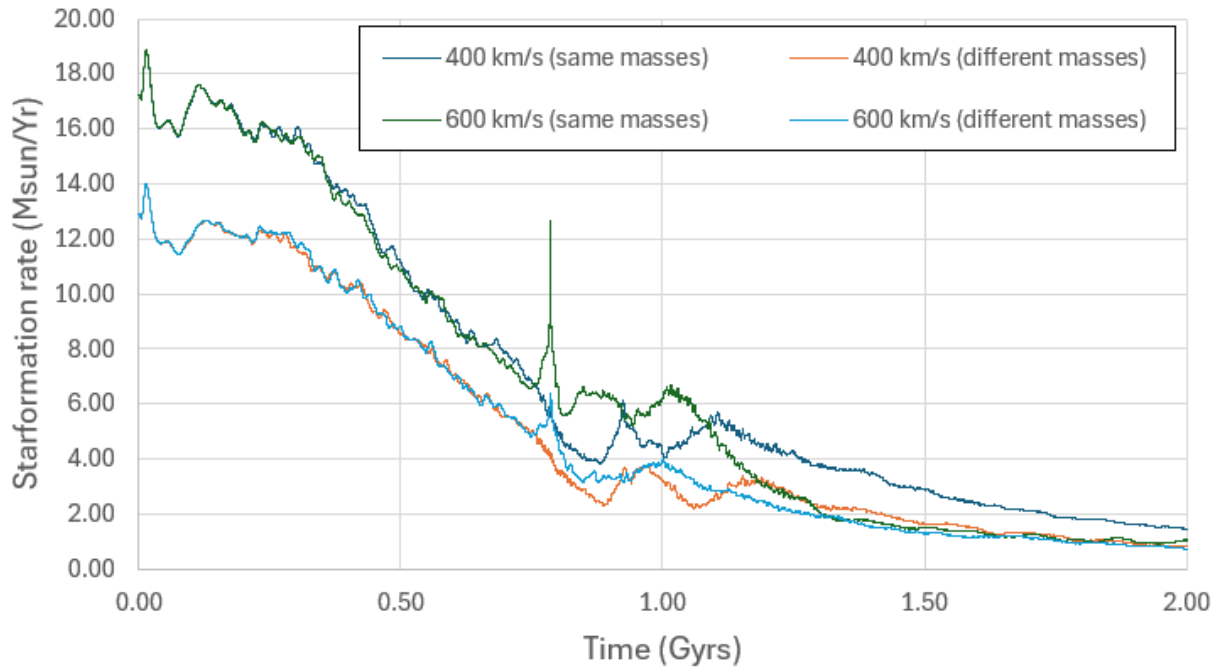

Figure 17 – Comparison between star formation rates in head-on collisions of identical and different galaxies, when no merger occurs.

Finally, in Figures 16 and 17, it is observed that the star formation rates gradually decline until, after a long period, they reach an approximately constant value.

**3.3 – Characteristics of head-on collisions between galaxies**

It is observed that, initially, all systems exhibit a very high rate of star formation, on the order of several tens of solar masses per year. This occurs due to the initial condition of the modeled galaxy, with its gas proportions being oversized relative to the amount of matter. However, it is noted that eventually this high star formation rate decreases to a lower and constant level, generally a few solar masses per year, which indicates a more balanced situation between the gas and matter proportions in the galaxies [15].

Another common feature observed is that whenever the visual interface records a collision of the galaxies, a peak in the system's star formation rate is also recorded, due to the interactions of the galaxies' gaseous disks [2].

Specifically in collisions with higher relative velocities, a large dispersion of the galaxies' gas is observed after the collisions, which is an expected phenomenon [3] and resembles the "splash bridges" [22]. In such situations, there is a drastic reduction in the star formation rates after the collision [2]. In our work, for an initial relative distance of 200kpc, the observed threshold is v = 200km/s.

On the other hand, galaxy collisions with lower relative velocities (i.e., velocities smaller than the threshold for a fixed relative distance of 200kpc) may lead to the merger of the galaxies, obviously depending on their masses. In these cases, several sets of peaks in the star formation rate are observed, with the first peak always associated with the first collision, when the galaxies pass through each other for the first time. Due to the gravitational capture of the galaxy system, these galaxies eventually interact again after some time, and a second peak in the star formation rate occurs when they pass through each other once more. This process repeats itself; the relative velocities of the galaxies decrease in each interaction, and the merger process occurs [2]. Astronomical observations of colliding galactic systems show that those with relative velocities between 100 km/s and 200 km/s are the most frequent [23].

Finally, the importance of the initial distance, or the distance of the first astronomical observation, between galaxies in the collision process is also highlighted. In our study, for galaxies with initial relative velocities of 300 km/s, depending on the masses of the galaxies and the initial observation distances, they may or may not merge.

However, there are also differences between the two case studies: collisions of two large galaxies and collisions of large-small galaxies. Larger peaks in star formation rates have been observed and are associated with galaxies that contain larger amounts of gas [2].

## 4 – Conclusion

The purpose of this study is to observe how changes in the initial parametrization of a system of two galaxies on a head-on collision course affect the star formation rate and the evolution of this system.

In the simulations with lower initial relative velocities (100 km/s to 200 km/s), a situation more common in observational studies [32], the merger of the two galaxies involved in the collision was observed. On the other hand, when the relative speed of galaxies is high, for example, on the order of 1000 km/s, no merger occurs after the collision; there is a large dispersion of gas from the galactic disks into the intergalactic medium and, consequently, a significant drop in the rate of star formation in these cases.

**Acknowledgments**: The author Gustavo N. Pereira received support from a PICME–OBMEP/CNPq Undergraduate Research Scholarship. We appreciate the suggestions and criticisms made by Dr Thiago S. Pereira and Dr Rubens E. G. Machado.